\documentclass{llncs}
\usepackage{makeidx}  
\usepackage{graphicx}
\begin{document}
\graphicspath{{./figs/}{./}}
\frontmatter          
\pagestyle{headings}  

\mainmatter              
\title{Using Linguistic Features to Estimate Suicide Probability of Chinese Microblog Users}

\author{
Lei Zhang\inst{1,2} \and
Xiaolei Huang\inst{3} \and
Tianli Liu\inst{4} \and
Zhenxiang Chen\inst{*2} \and
Tingshao Zhu\inst{*1,5}
}
\institute{
Institute of Psychology, Chinese Academy of Sciences (CAS), Beijing 100101, China
\and
University of Jinan, Shandong 250022, China
\and
China Networking Information Center, Chinese Academy of Sciences, China
\and
Institute of Population Research, Peking University, China
\and
Key Lab of Intelligent Information Processing, Institute of Computing Technology, CAS, China
}

\maketitle

\begin{abstract}
If people with high risk of suicide can be identified through social media like microblog,
it is possible to implement an active intervention system to save their lives.
Based on this motivation,
the current study administered the Suicide Probability Scale(SPS) to 1041 weibo users at Sina Weibo,
which is a leading microblog service provider in China.
Two NLP (Natural Language Processing) methods, the Chinese edition of Linguistic Inquiry and Word Count (LIWC) lexicon and Latent Dirichlet Allocation (LDA),
are used to extract linguistic features from the Sina Weibo data.
We trained predicting models by machine learning algorithm based on these two types of features,
to estimate suicide probability based on linguistic features.
The experiment results indicate that LDA can find topics that relate to suicide probability,
and improve the performance of prediction.
Our study adds value in prediction of suicidal probability of social network users with their behaviors.
\end{abstract}

\section{Introduction}

Along with the wide spread of Sina Weibo,
the leading microblog website in China,
more and more people are willing to express their thoughts and emotions on the Internet.
The messages they post have the potential to reveal their psychological indexes like suicidal ideation with or without intension.
In this paper we proposed a task that predict the suicidal ideation of Sina Weibo users.
By doing that, the group with high probability to commit suicide can be identified and lives may be saved by intervention.

First, we collect information both on suicidal ideation and messages of Sina Weibo users.
Suicidal ideation is measured by Suicide Probability Scale(SPS), a widely used psychological scale.
Messages are downloaded with a crawler.
Second,
Linguistic Inquiry and Word Count (LIWC) and Latent Dirichlet Allocation (LDA) are applied to messages to extract features.
Then we build a supervised machine learning model to predict SPS score.
The major contributions we made are:

\begin{itemize}
 \item
 This is the first large scale suicide probability detection study on social media users.
 \item
 Consistent with findings of previous studies, our study results suggested that topics extracted from texts can reveal more information than LIWC lexicon.
 \item We compared the predictive power of trained topics with inferred topics.
 The inferred topics have the equal, even higher, predictive power than trained topics. On the other hand, inferred topics are more reusable than trained topics.
\end{itemize}

\section{Related Work}

\subsection{Sensing Suicidal Ideation}

Hotline service and face-to-face diagnose are two effective methods for suicide intervention, but both have limitations in lacking of initiative.
Mental health professionals are encouraged to contact with the high-risk group of suicide users, but with low efficiency.

Enlightened by the phenomenon that many people post suicide notes on the Internet, online suicidal ideation sensation shows out an effective way of proactive suicide defense.
Even though it is not yet clear to what extent suicide notes on social networks produce bad influence or even actually induce copycat suicides \cite{ruder2011suicide}, the suicide intervention of network users is a promising way to save lives.
To our knowledge, there are only two studies which are focused on detecting suicidal ideation on the Internet.
Silenzio proposed a method to identify hard-to-reach population who reported high rates of suicide ideation \cite{silenzio2009connecting}.
Jared et al. made a more straightforward progress to create a message filter on Twitter using words and phrases created from suicide risk factors \cite{jashinsky2013tracking}.

In addition, there are some similar studies which sense negative emotions related to psychopathy or mental health:
Wald predict Twitter users score in the top 1.4 percent on psychopathy \cite{wald2012using} using public visible information on Twitter.
Although the accuracy several algorithms achieved are not good enough to replace diagnosis, still the research can play the effect of early warning.

\subsection{Predict Psychological Indexes with Social Network}

In recent years, it becomes a research focus using social network data to predict psychological indexes automatically. Psychological indexes that had been well studied included personality, mental health, subjective well-being and so on.

Golbeck et al. build predictive models on both Twitter and Facebook
\cite{Golbeck2011Twitter,Golbeck:2011:PPS:1979742.1979614} .The information they collected from two websites involves language features, personal info, activities and preferences, internal website statistics, composite features, structural and so on. Then Sibel and Golbeck proposed a comparative study about predicting personality with social behavior \cite{adali2014predicting}. The latest progress of theirs is that with the follower connections in the Twitter network, they computed the political preferences of an organization's Twitter followers \cite{Golbeck2014177}.

A solid predict model requires large scale data, that's why Kosinski et al. made such a great effort that they collected more than 6,000,000 test results together with more than 4,000,000 individual
Facebook profiles. With the help of large scale data, they create a Users-Likes matrix to predict a wide range of people's personal attributes, ranging from sexual orientation to Big-Five personality
\cite{Kosinski09042013}. Quercia map the psycho-score from Facebook to Twitter and create a predictive model on Twitter \cite{Quercia2011Twitter}. Schwartz et al. extracted words, phrases and topics from the same dataset and proposed the open-vocabulary approach to predict age, gender and personality \cite{10.1371/journal.pone.0073791}.

Social networks in non-English speaking country are also proved to have the predictive power to psychological indexes.
Bai predict personality on both RenRen and Sina Weibo \cite{bai2012determining,Bai2013weibo}. Hao predict subjective well-being on Sina Weibo \cite{hao2014sensing}.

\subsection{Linguistic Feature Extraction}

Dictionary based methods are generally used by psychologists to analyze people's linguistic differences.
For example, a positive emotion words dictionary might contains words like ``happy'', ``good'', ``love'', ``nice'' and ``sweet''.
Counting the frequency of positive emotion words used by introverts and extroverts provides psychologists a quantitative result that distinguish two kinds of people on their linguistic differences.
Linguistic Inquiry and Word Count (LIWC) is a hand-crafted dictionary employed in social psychology studies which has been widely used \cite{pennebaker2001linguistic}.
Consider ``positive emotion'' is only one category in our example, the 2007 version of LIWC provides 64 categories to cover a wide range of human linguistic features.
So for every document we are interested in, LIWC can provide 64 features to analysis for researchers.

Topic models are a series of algorithms to uncover the salient information lay behind document collections.
Among these algorithms, the unsupervised algorithm Latent Dirichlet Allocation (LDA) which proposed by David Blei on 2003 \cite{Blei:2003:LDA:944919.944937} made topic models even more well known.
The results of LDA are two matrixes, one is document-topic matrix and the other topic-term matrix.
People can get to know the possible meaning of each topic by viewing the top ranked words estimated for the topic in the topic-term matrix.
Just like the 64 features, the document-topic matrix provides similar features only without category names (number of topics can be seen as category name but it is not intuitively meaningful).
Topic models can help us developing new ways to summarize large archives of texts and predicting specific features of users or the messages themselves.

\section{Methods}

\subsection{Data Collection}

Sina Weibo is the biggest microblog website in China. During May and
Jun 2014, 1038 Sina Weibo users are recruited to complete Suicide
Probability Scale (SPS) \cite{cull1982suicide}.
The SPS consists of four sub-scales with eight items in each sub-scale.
The final score of SPS is the sum of all the four parts.

\begin{figure*}
  \includegraphics[width=1\textwidth]{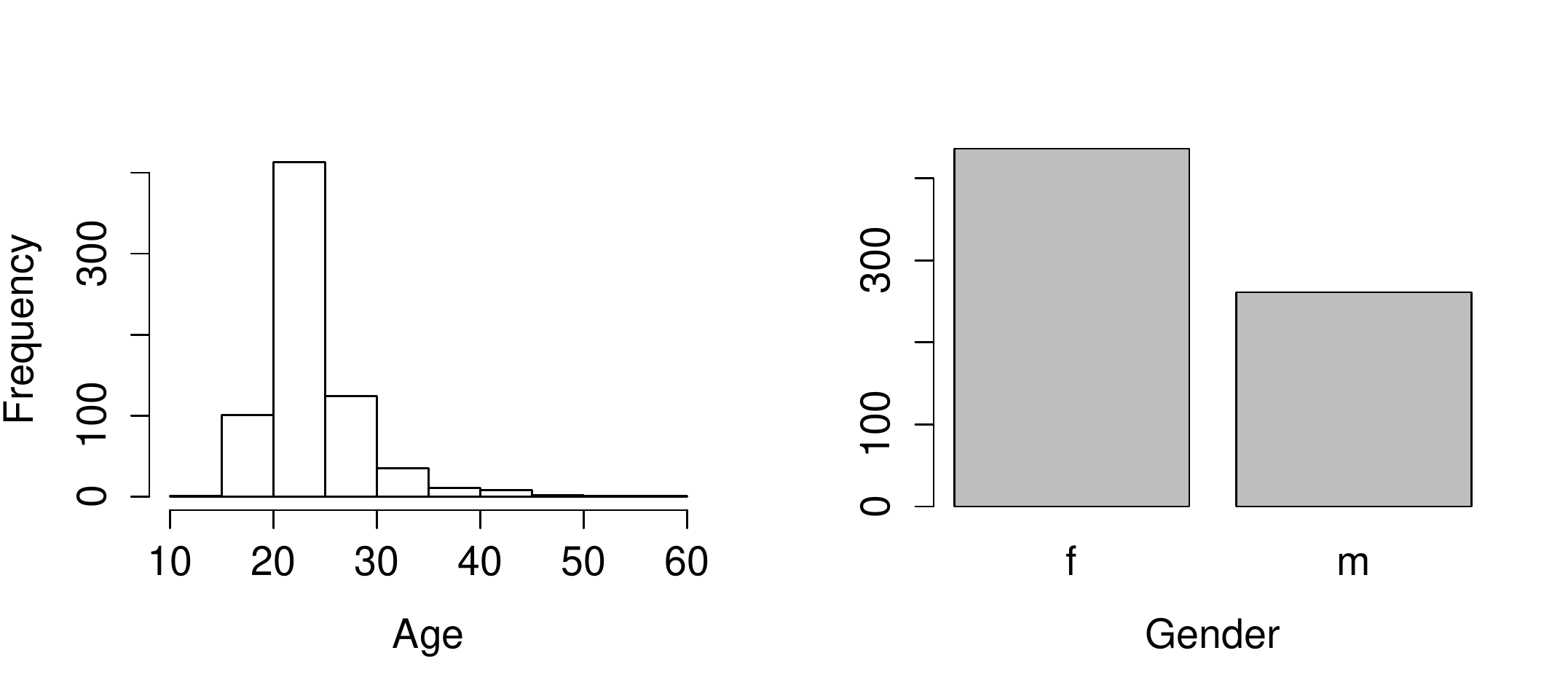}
  \caption{\label{fig:age} Distribution of age and gender}
\end{figure*}

The participants who complete the scale can get paid for 30 Chinese yuan.
The age of participants range from 14 to 63 and the average value is $24$.
The gender of participants is not evenly distributed that in totally 647 women and 391 men involved in the test.
Figure\ref{fig:age} shows age and gender distribution of all the participants.

\begin{figure*}
  \centering \includegraphics[width=0.6\textwidth]{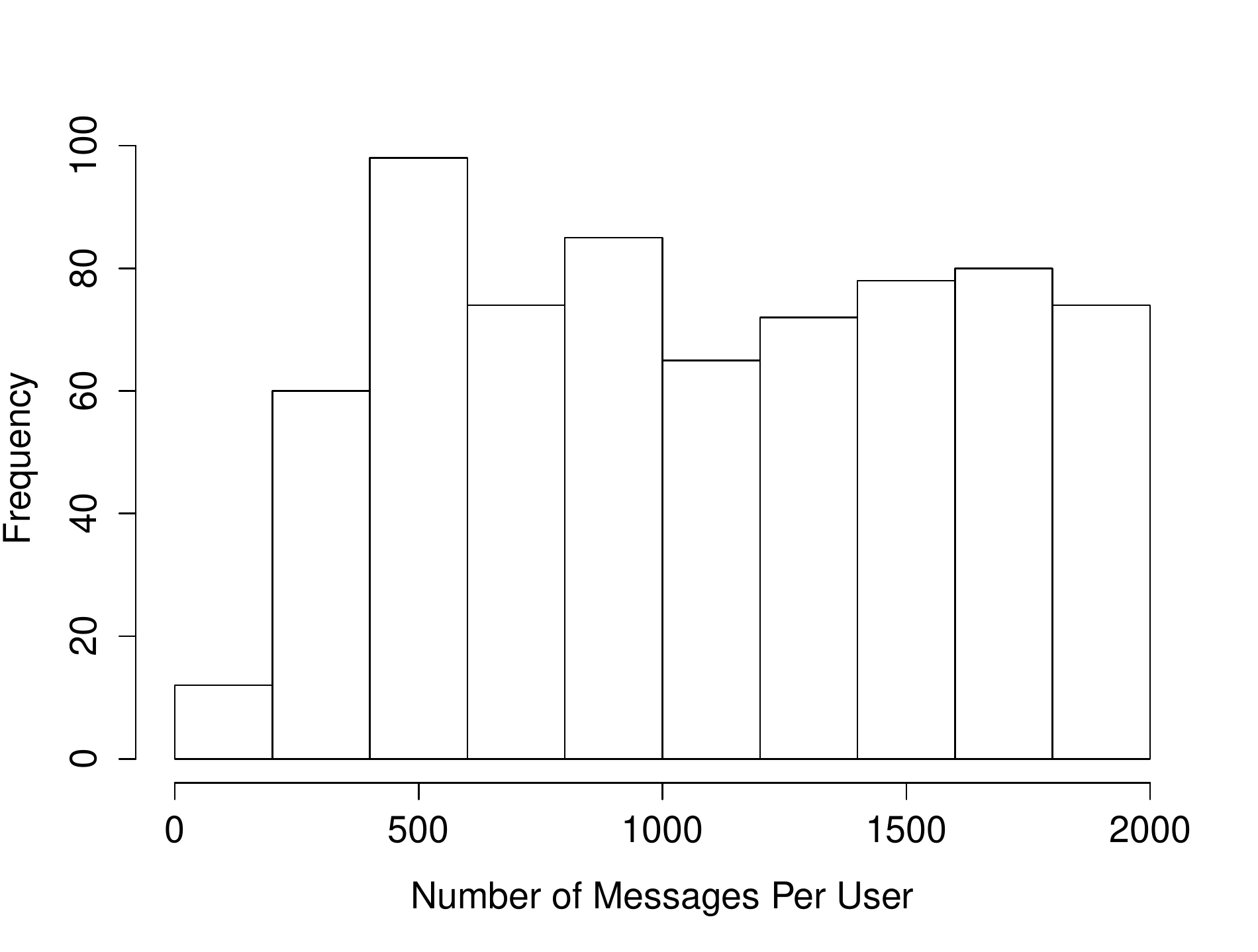}
  \caption{\label{fig:textcount} Distribution of the number of messages}
\end{figure*}

To associate the score to Sina Weibo users, we create a crawler program to download participants' recent 2000 messages on the Sina Weibo (download all the messages if there are less than 2000 messages). Figure \ref{fig:textcount} shows the distribution of the number of messages per user.

To ensure the reliability of SPS score, we excluded the users who answered one question using less than 2 seconds or more than 10 seconds.
Users with less than 20KB texts were also excluded, because a small amount of text are not likely reveal psycho-information of users.
Our final analytic sample included 697 Sina Weibo Users.

\subsection{Text Process}

The length of messages on microblog are always restricted.
Sina Weibo allow users post less than 140 Chinese characters each time.
Even so, Sina Weibo is a very
good source of text for web content mining because all the messages are publicly visible and the amount is huge.
LWIC and topic models are used widely on Facebook status updates and Twitter messages which are all written in English.
On Chinese corpus like Sina Weibo messages,
the process details and their performance remains questions.

\textbf{Preprocess}. Firstly,
we removed some Sina Weibo specific marks (like retweet marks, hashtags) and the words don't speak by the author (like retweets, replies).
Then a segmentor is used \cite{che2010ltp}, as Chinese words are not separated by default.
The segment model of the segmentor is trained by manual segmented Sina Weibo sentences.
After segmenting, we removed the stop words and single words (consider the express unit of Chinese).

\textbf{LIWC}. The dictionary we used is a revised and improved version
of the original LIWC which fit for simplified Chinese \cite{gao2013developing}.
This simplified Chinese version extended the existing 64 categories to 88 categories.
To make the LIWC features normalized, the counts of every categories are divided by the document length for each document.

\textbf{Topic Model}. We use an implementation of LDA algorithm provided by the Mallet toolkit.
Aggregating all the messages from a user as a single document allows us to compute topics at user level.
It can be considered as an application of the author-topic model \cite{steyvers2004probabilistic}.
Different from LIWC, we remove all the single words from segmented texts before topic modelling because single word is not the basic ideographic unit in
Chinese.

There are two strategies to get the documents-topics matrix.
The first strategy is simply train all the 697 documents and use the post-estimated probability as features.
As shown in figure \ref{fig:ldaPre}, from 10 to 300, a large range of number of topics are empirically used.
All other parameters are set as their default.
The second strategy is to train a model with another corpus then infer the 697 documents.
Consider the trades of suicidal ideation may be rare, the chance to estimate a topics about suicidal ideation may be little.
To maximum this chance on our own minds, we trained 12 models (with the same number of topics of the first strategy) which only include texts from high suicidal ideation participants.
The high suicidal ideation group involves 107 participants whose SPS scores are higher than the sum of the mean value and one standard deviation.
Additionally appended to the corpus, there are 30 more documents from Sina Weibo users that are confirmed to have committed suicide.
At last, the topics of texts from 697 users with SPS score are inferred with the pre-trained model trained by 137 documents.

\subsection{Predict Model}

We use caret package in R to conduct linear regression analysis with stepwise selection \cite{caret}.
In building predict model, text features (LIWC features and topics features) are independent variable and SPS score is dependent variable.
Caret creates a unified interface for modeling and prediction which contains a variety of ``helper'' functions and classes. The metric we use to evaluation uncertainty of predicted score is RMSE (Root Mean Square Error) \cite{zhu2012evaluation}.

\section{Results}

With the help of caret package,
a 10-fold cross validation is used to take most advantage of data and avoid potential over-fitting.

\subsection{Predictive Power of LIWC}

A correlation analysis is run for each language feature of LIWC.
The dimension with highest correlation coefficient of $0.13$, is ``inhibition'' words (e.g., ``block'', ``constrain'', ``stop'' in English).
The second highest correlated dimension is ``cognitive processes''
(e.g., ``cause'', ``know'', ``ought'' in English), with the correlation coefficient of $0.09$.
Both of the correlations are significant at $p < 0.01$ level.

Because we have only one parameter to predict, its RMSE is $15.43$.
Using this number as the baseline value, we can compare it with results topics model obtained.

\subsection{Predictive Power of Topic Model}

As is discussed in chapter 3.2, we run topic features in two different way.

\begin{figure}
  \centering \includegraphics[width=0.6\textwidth]{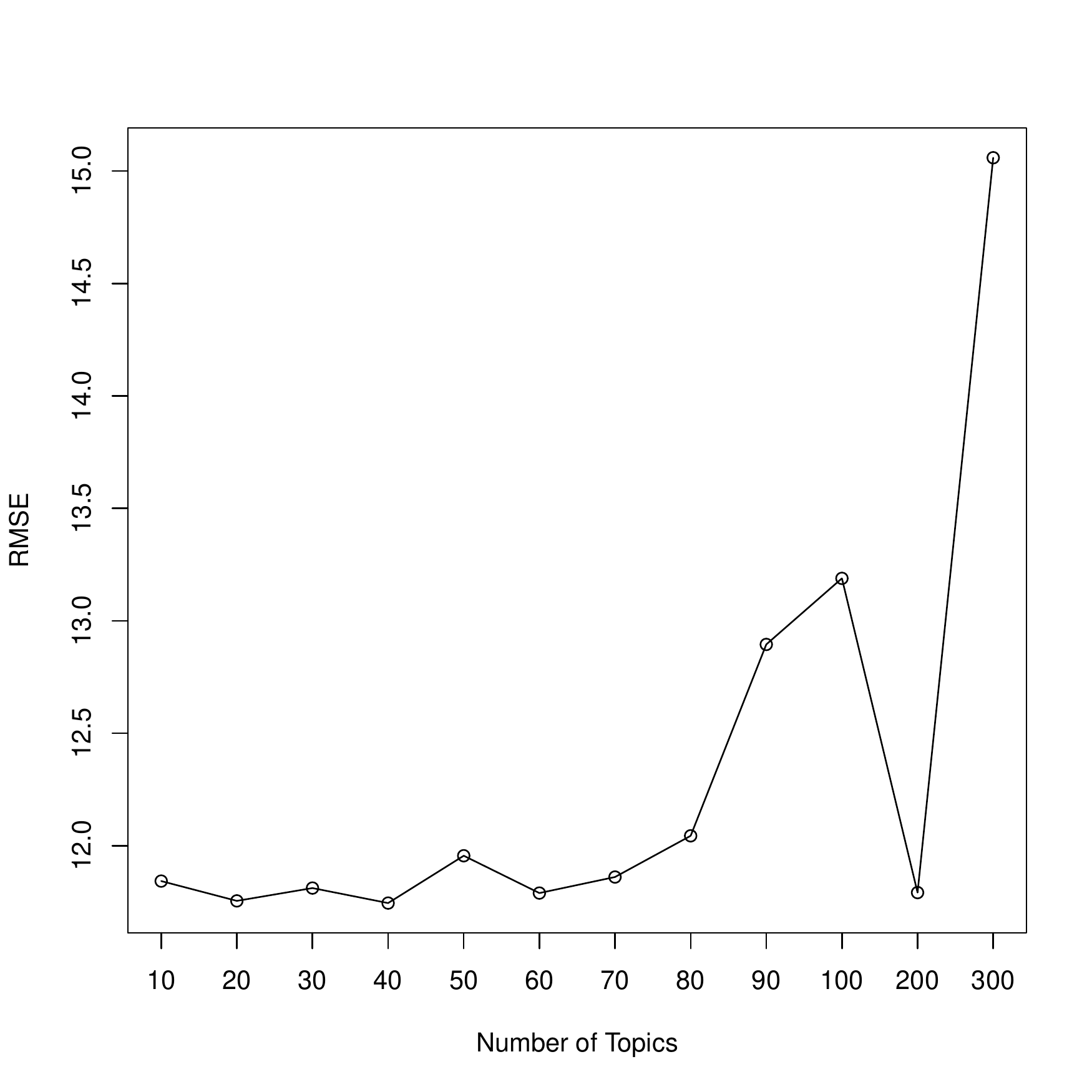}
  \caption{\label{fig:ldaPre} RMSE for SPS using trained topics as features across number of topics}
\end{figure}

\textbf{Result of trained topics}  As is shown in figure
\ref{fig:ldaPre} we experimented with LDA using 10, 20, ...,  100, 200 and 300 topics. When the number of topics is no more than 70, RMSE fluctuated only slightly (from $11.68$ to $11.97$). When a bigger number is set to LDA, RMSE increased obviously (top to $15.06$).
It indicates that when the granularity of topics are too small, the topics are more likely fail to capture the clue of suicidal ideation.
The best performed RMSE is $11.68$ when number of topics is set to 70, which is much better than the result of LIWC ($15.43$).
It also confirmed the earlier research that topic features outperform LIWC features on predictive power \cite{resnik-garron-resnik:2013:EMNLP,10.1371/journal.pone.0073791}

\begin{figure}
  \centering \includegraphics[width=0.6\textwidth]{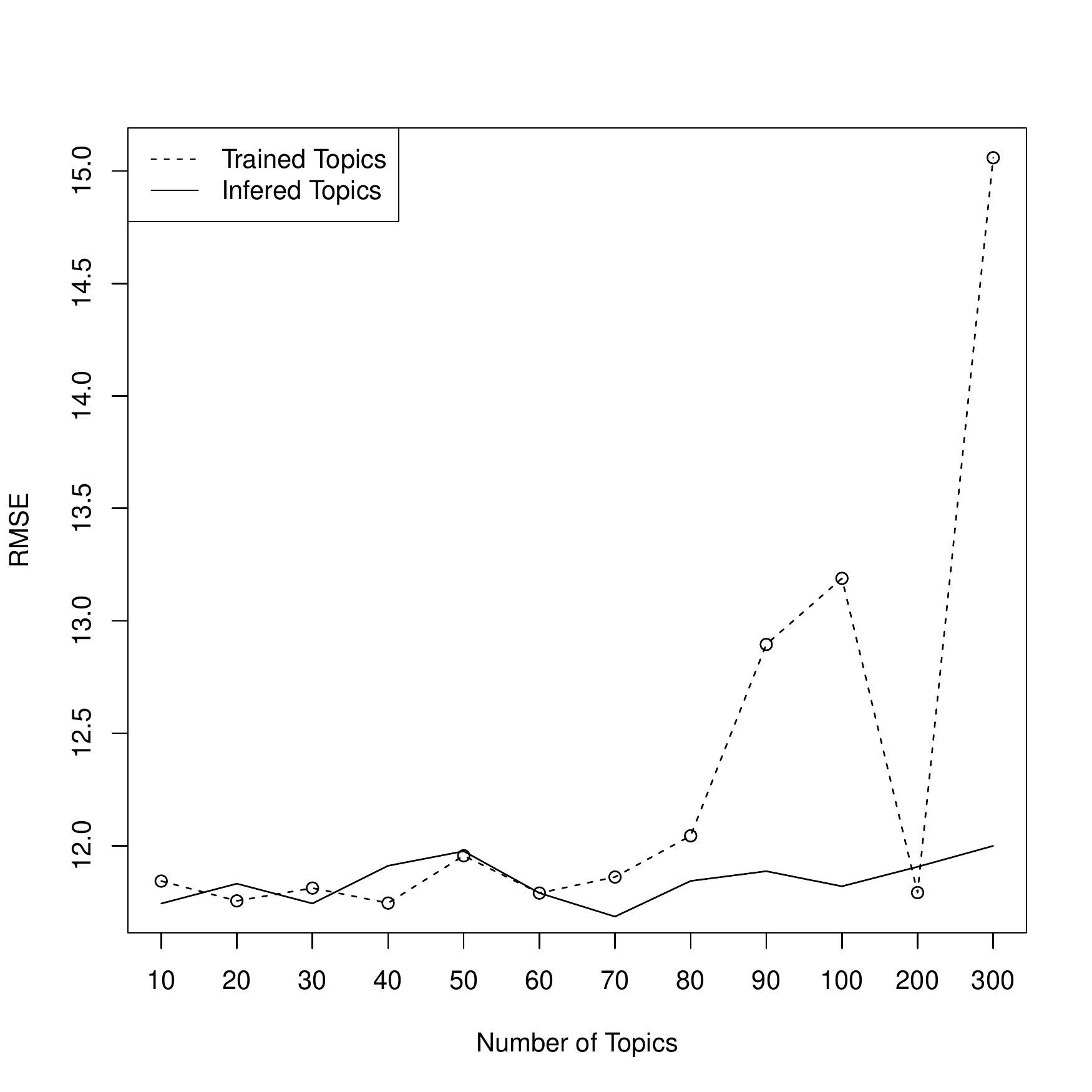}
  \caption{\label{fig:ldaCompare} RMSE of for SPS across number of topics. Inferred topic features compared with trained topic features}
\end{figure}

\textbf{Result of inferred topics}
As presented in Figure\ref{fig:ldaCompare}, we plot RMSE with the same number of topics set by trained topics strategy.
When the number of topics is less than 70,
RMSE fluctuated slightly and performed almost equally with trained topics.
On a bigger number of topics, inferred topics performed still steadily which was different from trained topics.
The reason may be the small training set with high suicidal ideation is easier to estimate topics related to suicidal ideation.

\begin{figure}
  \centering \includegraphics[width=0.6\textwidth]{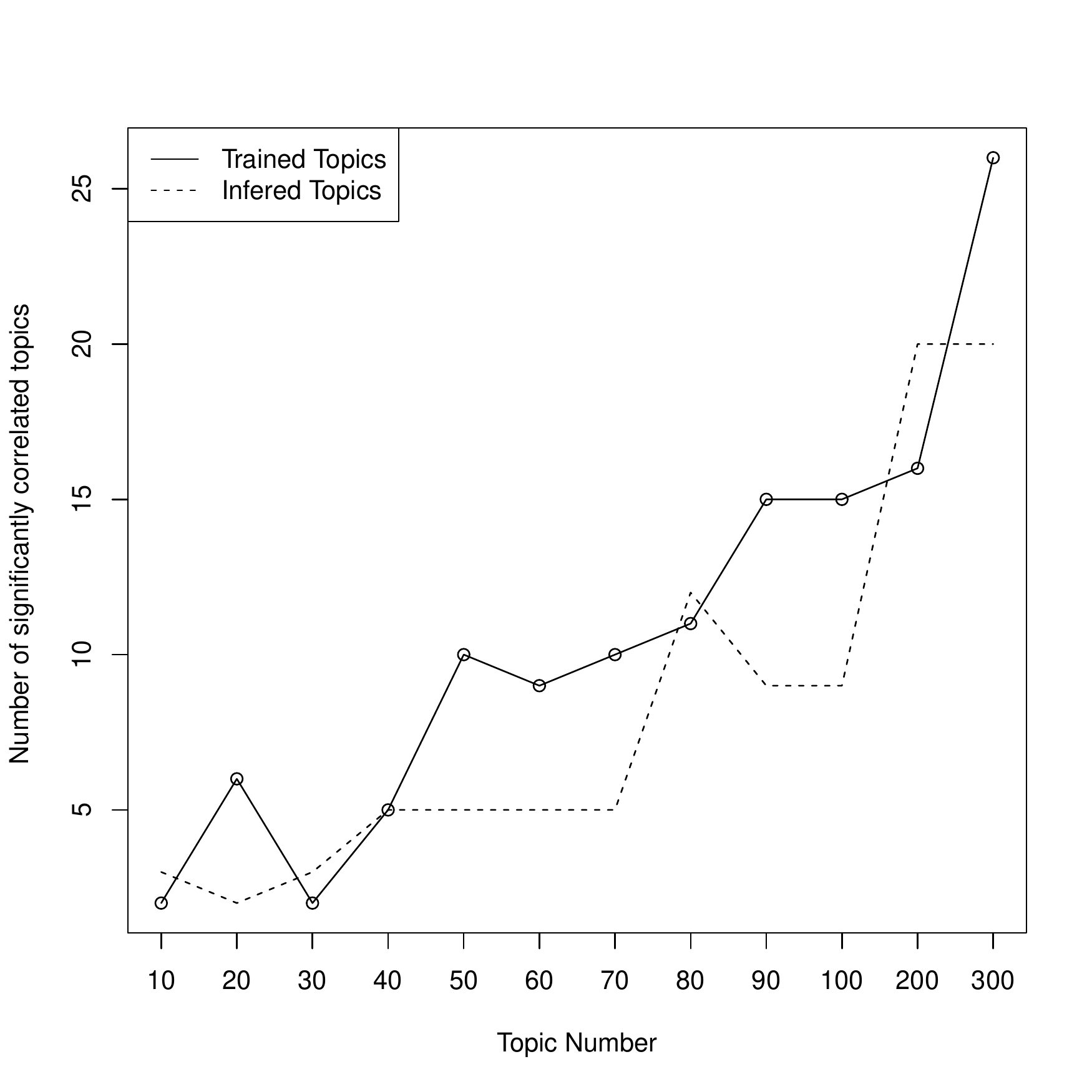}
  \caption{\label{fig:corrNum} Number of significantly correlated topics}
\end{figure}

We also inspect the number of significantly correlated topics and the maximum correlation coefficient of single topic.
As shown in figure \ref{fig:corrNum}, along with the increasing of number of topics, the number of significantly correlated topics also increase but the maximum of correlation coefficient didn't show in the models with highest number of topics.
We are also curious about how to estimate a topic which have correlation as high as possible.
From table \ref{tb:maxcc}, it seems a suicide correlated topic has nothing to do with number of topics and whether it's trained or inferred.

\begin{table}[ht]
\centering
\caption{\label{tb:maxcc} Maximum correlation coefficient of single topic}
\begin{tabular}{rrr}
  \hline
Number & Trained & Inferred \\
  \hline
10 & 0.11 & 0.09 \\
  20 & 0.11 & 0.14 \\
  30 & 0.15 & 0.10 \\
  40 & 0.11 & 0.14 \\
  50 & 0.13 & 0.11 \\
  60 & 0.15 & 0.14 \\
  70 & 0.11 & 0.13 \\
  80 & 0.14 & 0.13 \\
  90 & 0.11 & 0.14 \\
  100 & 0.14 & 0.12 \\
  200 & 0.15 & 0.14 \\
  300 & 0.15 & 0.12 \\
   \hline
\end{tabular}
\end{table}

\subsection{Predictive Power of LIWC+Topic Model}

As is shown in Table \ref{tb:rmse}, we see that LIWC didn't add much value on predictive power for SPS score.
When we combine LIWC and trained topics as features, predict model was improved with regard to big number of topics compared with using trained topics only.
When we combine LIWC and inferred topics as features, predict model was not improved. In general, only inferred topics can achieve highest performance no matter with or without help from LIWC.

\begin{table}[ht]
\centering
\caption{\label{tb:rmse} Comparison of LIWC and topic features within predictive models of SPS score using RMSE as metric }
\begin{tabular}{rrrrr}
  \hline
 Number & Trained & Inferred & LIWC+Trained & LIWC+Inferred \\
  \hline
10 & 11.84 & 11.74 & 11.79 & 11.83 \\
  20 & 11.75 & 11.83 & 11.79 & 11.79 \\
  30 & 11.81 & 11.74 & 11.78 & 11.84 \\
  40 & 11.74 & 11.91 & 11.76 & 11.86 \\
  50 & 11.96 & 11.98 & 11.82 & 11.79 \\
  60 & 11.79 & 11.79 & 11.80 & 11.86 \\
  70 & 11.86 & 11.68 & 11.88 & 11.85 \\
  80 & 12.04 & 11.84 & 11.79 & 11.82 \\
  90 & 12.90 & 11.89 & 11.80 & 11.80 \\
  100 & 13.19 & 11.82 & 11.80 & 11.84 \\
  200 & 11.79 & 11.91 & 11.80 & 11.88 \\
  300 & 15.06 & 12.00 & 11.82 & 11.80 \\
   \hline
\end{tabular}
\end{table}

\section{Conclusion}

In this paper, we established models to detect Sina Weibo users' suicidal ideation using two different NLP methods.
The best RMSE value achieved with stepwise linear regression is around 11 at 1-32 scale.
The study results suggested that suicidal ideation is predictable.
But the predictive accuracy need to be improved.
Furthermore, as trades of suicidal ideation may be rare, innovatively we attempt topics inferred from a model trained by high suicidal ideation group.
The inferred topics reached equal or higher quota in the predict task.
The same method is worth trying on other predict tasks.
Moreover, some conclusions made by previous works are confirmed in our study.

\section{Acknowledgments}
The authors gratefully acknowledge the generous support from National High-tech R\&D Program of China (2013AA01A606),
National Basic Research Program of China(973 Program£¬ 2014CB744600),
Key Research Program of CAS (KJZD-EW-L04) and Strategic Priority Research Program (XDA06030800) from Chinese Academy of Sciences.

\bibliographystyle{splncs03}
\bibliography{icpca}

\end{document}